\documentclass[prl,twocolumn,floats,showpacs,superscriptaddress]{revtex4}
\usepackage{psfig}
\usepackage{amsmath,amssymb,amsthm}
\usepackage{graphicx}
\usepackage{psfrag}
\usepackage{bm}

\newcommand{\lam}{\lambda}
\newcommand{\sig}{\sigma}
\newcommand{\al}{\alpha}
\begin{document}

\title{Entangled networks, synchronization and optimal 
network topology}

\author{Luca Donetti} 
\affiliation{
  Departamento de Electromagnetismo y F{\'\i}sica de la Materia and 
  Instituto  Carlos I de F{\'\i}sica Te{\'o}rica y Computacional \\ Facultad de
  Ciencias, Universidad de Granada, 18071 Granada, Spain}
\author{Pablo I. Hurtado}
\affiliation{Department of Physics, Boston University, Boston, Massachusetts 
02215, USA}
\author{Miguel A. Mu{\~n}oz} 
\affiliation{
  Departamento de Electromagnetismo y F{\'\i}sica de la Materia and 
  Instituto  Carlos I de F{\'\i}sica Te{\'o}rica y Computacional \\ Facultad de
  Ciencias, Universidad de Granada, 18071 Granada, Spain}
\date{\today} 
\begin{abstract} 

  A new family of graphs, {\it entangled networks}, with optimal
  properties in many respects, is introduced. By definition, their
  topology is such that optimizes synchronizability for many dynamical
  processes. These networks are shown to have an extremely homogeneous
  structure: degree, node-distance, betweenness, and loop
  distributions are all very narrow. Also, they are characterized by a
  very interwoven (entangled) structure with short average distances,
  large loops, and no well-defined community-structure. This family of
  nets exhibits an excellent performance with respect to other flow
  properties such as robustness against errors and attacks, minimal
  first-passage time of random walks, efficient communication, etc.
  These remarkable features convert entangled networks in a useful
  concept, optimal or almost-optimal in many senses, and with plenty
  of potential applications computer science or neuroscience.
\end{abstract} 

\maketitle 

The ubiquitous presence of networks in Nature and social sciences is
one of the main findings in the study of complex systems. The topology
of such networks has been profusely studied
\cite{Reviews} and some basic architectures have
been discovered. The {\it scale-free} one, characterized by a
power-law connectivity-distribution, is probably the most widely
studied and celebrated, while other examples are small-world,
hierarchical, Apollonian, static networks, etc \cite{Reviews}. Right
after the first topological studies, the interest has shifted to the
analysis of functional or dynamical aspects of processes occurring on
networks, the evolution of the network topology, and the interplay
between these last two dynamical features. Indeed, this ``network
perspective'' has become a new paradigmatic way to look at complex
systems. One particular issue that has attracted much interest because
of its conceptual relevance and practical implications is the study of
the {\it synchronizability} of individual dynamical processes
occurring at the vertices of a given network. How does
synchronizability depend upon network topology? This problem is much
more general than it seems at first sight, as it is directly related
to the question of how difficult it is to transmit information across
the net or how difficult is for the sites to ``talk'' to each
other. For example, a recently addressed important task is to
determine the most efficient topology for communication networks both
with and without traffic congestion \cite{Catalans}.
Other problems as the minimization of first-passage times of random
walkers on networks, the optimal topology in social networks to reach
consensus, or the performance optimization of Hopfield
neural-networks \cite{Torres,BJK} are also similar in essence. Hence,
the issue of synchronizability is linked to many specific problems in
different disciplines as computer science, biology, sociology,
etc. \cite{Pecora,Catalans}. Some aspects of these problems have been
already tackled; a key contribution is due to Barahona and Pecora (BP)
\cite{Pecora} who established a criterion based on spectral techniques
to determine the stability of synchronized states on networks.

The criterion is as follows. Consider a dynamical process $\dot{x}_i =
F(x_i) - \sig \sum_j L_{ij} H(x_j)$, where $x_i$ with $i \in
{1,2,... ,N}$ are dynamical variables, $F$ and $H$ are evolution
and coupling functions respectively, $\sigma$ is a constant, and
$L_{ij}$ is the Laplacian matrix, defined by $L_{ii}=k_i$ (the
connectivity degree of node $i$), $L_{ij}=-1$ if nodes $i$ and $j$ are
connected, and $L_{ij}=0$ otherwise. A standard linear stability
analysis can be performed by i) expanding around a synchronized state
$x_1=x_2=\ldots=x_N=x^s$ with $x^s$ solution of $\dot{x^s} = F(x^s)$,
ii) diagonalizing $L$ to find its $N$ eigenvalues $0=\lam_1 < \lam_2
\le \ldots \le \lam_N$, and iii) writing equations for the normal
modes $y_i$ of perturbations $ \dot{y}_i = \left[ F'(x^s) - \sig
\lam_i H'(x^s) \right] y_i $ which have all the same form but
different effective coupling $\al=\sig \lam_i$.  BP observed that the
maximum Lyapunov exponent is in general negative only within a bounded
interval $[\alpha_A,\alpha_B]$, and a decreasing (increasing) function
below (above) (see fig. 1 in
\cite{Pecora}). Requiring all effective couplings to lie within such an
interval, $ \al_A < \sig \lam_2 \le
\ldots \le \sig \lam_N < \al_B$, it is straightforward to conclude
that a synchronized state is linearly stable on a network if and only
if $ \frac{\lam_N}{\lam_2} < \frac{\alpha_B}{\alpha_A}.$ Notice that
the left hand side depends only on the network topology while the
right hand side depends exclusively on the dynamics (through $F$ and
$G$, and $x^s$). Moreover, the interval in which the
synchronized state is stable is larger for smaller eigenratios
$\lam_N/\lam_2$, whence one concludes that {\it a network exhibits
better synchronizability if the ratio $Q = \lam_N/\lam_2$ is as small
as possible}, independently of the dynamics.

This letter is devoted (i) to build-up networks with a fixed number of
nodes $N$ and average connectivity $ \langle k \rangle$, exhibiting a
degree of synchronizability as high as possible (i.e. minimizing $Q$),
(ii) to explore the topological features converting them into highly
synchronizable, and (iii) to highlight their connection to networks
optimizing other flow or connectivity properties relevant in
neuro-computing, computer science, or graph theory.

First, we overview how $Q$ behaves in some well-known topologies. For
networks with the small-world property \cite{Reviews} $Q$ is smaller
than for deterministic graphs or purely random networks
\cite{Pecora}. This was attributed to the existence of short
characteristic paths between sites. However, Nishikawa {\it et al.} in
a study of other small-world networks concluded that $Q$ decreases as
some heterogeneity measures decrease, even if the average distance
increases \cite{Nishikawa}. Also, Hong et al. concluded
that $Q$ decreases whenever the betweenness heterogeneity decreases
\cite{Hong}.
In order to extend and systematize these results and construct optimal
synchronizable networks, and in the absence of a better strategy, we
define a numerical algorithm able to minimize $Q$ and search for such
optimal nets.

Our optimization algorithm is a modified simulated annealing
initialized with a random network with $N$ nodes and 
an average connectivity-degree $ \langle k \rangle$. At each step the
number of rewiring trials is randomly extracted from an exponential
distribution. Attempted rewirings are (i) rejected if the updated
network is disconnected, and otherwise (ii) accepted if $\delta Q=
Q_{final}-Q_{initial} <0$, or (iii) accepted with probability
\cite{Penna}
$p = \min\left( 1, [1-(1-q) \delta Q/T]^{1/(1-q)}\right)$ (where $T$
is a temperature-like parameter) if $\delta Q \geq 0$. In the $q \to
1$ limit the usual Metropolis algorithm is recovered, while we choose
$q=-3$ as it gives the fastest convergence (though results do not
depend on this, as already verified in \cite{Penna}). The first $N$
rewirings are performed at $T=\infty$, and they are used to calculate
a new $T$ such that the largest $\delta Q$ among the first $N$ ones
would be accepted with large probability; in particular, we take
$T=(1-q)\cdot(\delta Q)_{max}$. $T$ is kept fixed for $100N$ rewiring
trials or $10N$ accepted ones, whichever occurs first. Then, $T$ is
decreased by $10\%$ and the process iterated until there is no change
during $5$ successive temperature steps, assuming that a (relative)
minimum of $Q$ has been found. Most of these details can be changed
without affecting significatively the final results, while the main
drawback of the algorithm is that the calculation of eigenvalues is
slow.

\begin{figure}[tbp]
\centering \includegraphics[width=.65 \linewidth]{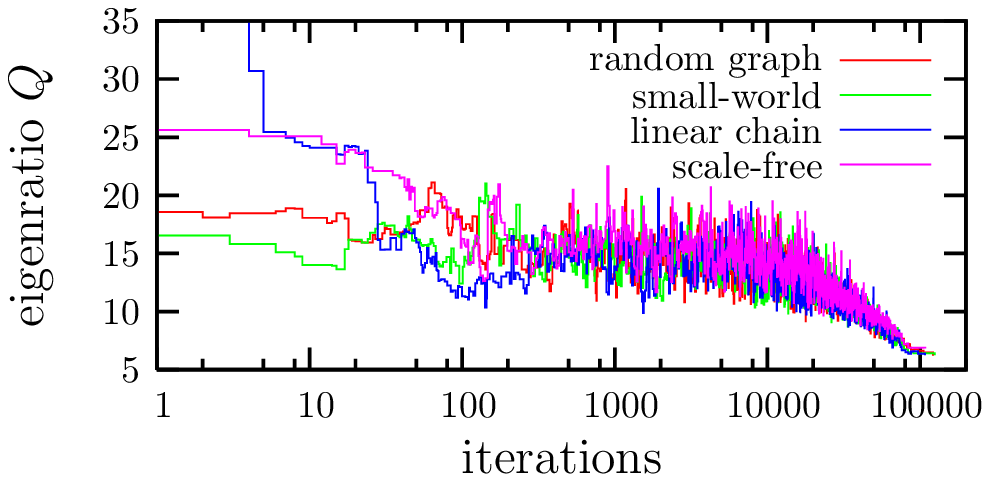} \centering~~
 \includegraphics[width=.3 \linewidth]{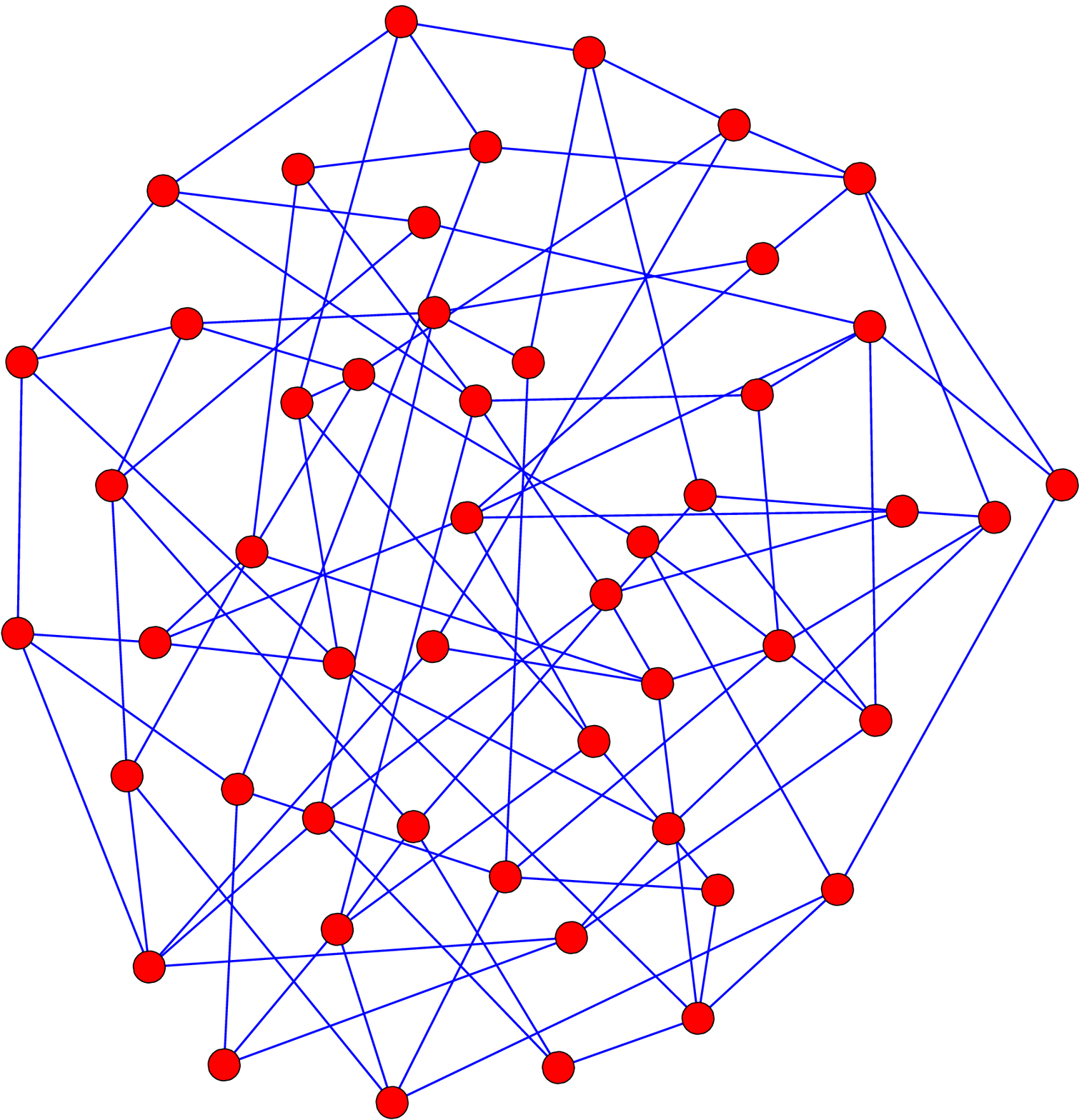}
\vspace{-0.3cm}
\caption{Eigenvalue ratio, $Q$ as a function of 
the number of algorithmic iterations. Starting from different initial
conditions, with $N=50$, and $\langle k \rangle=4$, the algorithm
converges to networks, as the depicted one, with very similar values
of $Q$.}
\label{Fig1}
\end{figure}

\begin{figure}
\centerline{ 
\psfig{file=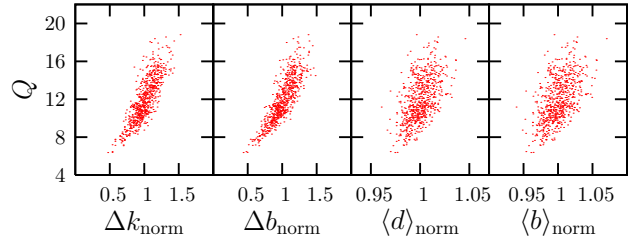,width=8.2cm}}
\vspace{-0.3cm}
\caption{ 
Relation between the ratio $Q$ and (i) node-connectivity standard
deviation, (ii) betweenness standard deviation, (iii) average
node-distance, and (iv) average betweenness. The subscript ``norm''
stands for normalization with respect to the respective mean-values,
centering all the measured quantities around $1$.}
\label{Fig2}
\end{figure}

The network found by different runs of the algorithm is unique (in
most of the cases) as long as $N$ is small enough ($N \lesssim 30$),
while they are slightly different if $N$ is larger ($N=2000$ is the
larger size we optimized). This indicates that the
eigenvalue-ratio absolute minimum is not always found, and that the
evolving network can remain trapped in some ``metastable''
state. Nevertheless, the final values of $Q$ are very similar from run
to run as shown in fig.~\ref{Fig1}. This fact makes us confident that
a reasonably good and robust approximation to the optimal topology is
obtained in general, though, strictly speaking, we cannot guarantee
that the optimal solution has been actually found.
To gain some insight into the topological traits favoring a small $Q$,
we measure some quantities during the evolution and plot them versus
the changing eigenratio. It turns out (as shown in fig.~\ref{Fig2})
that there is a strong correlation between the tendency of $Q$ to
decrease and an increase in the homogeneity (lowering variances) of
the degree, average-distance and betweenness distributions. In a
nutshell, the more synchronizable the network the more homogeneous it
is. Also, the average distance and betweenness tend to diminish with
$Q$, though these quantities are much less sensitive that their
corresponding standard deviations (fig.~\ref{Fig2}).
The emerging narrow betweenness distribution is in sharp contrast with
that of networks with a strong community structure
\cite{newman1}. Indeed, a well known method to detect communities
consists in removing progressively links with the largest betweenness
\cite{newman1}. The method leads to sound results whenever the
betweenness is broadly distributed. Hence, well-defined communities do
not exist in the emerging optimal net.

Further inspection of these networks reveals another significant
trait: the absence of short loops. This can be quantified by the {\it
girth} (length of the shortest loop) or more accurately by the
 average length, $\langle \ell
\rangle$, of the
shortest loop passing through each node. In general, the clustering
coefficient vanishes, as loops are larger than triangles. Indeed, for
small values of $N$ and $k$, it is possible to identify the resulting
optimized networks, as they have been studied in the mathematical
literature: some of them are {\it cage graphs}. Let us recall that a
$(k, g)$-cage graph is a $k$-regular graph (i.e. with a delta-peaked
connectivity distribution) of girth $g$ having the minimum possible
number of nodes. For $k=3$ and $N=10, 14$, and $24$, respectively, the
optimal nets found by the algorithm are cage-graphs with girth $5$,
$6$, and $7$ (called $Petersen$, $Heawood$ and $McGee$ graphs)
respectively (see fig.~3 and \cite{Wolfram}). For other values of $N$
cage graphs do not exist but, in all cases, networks with very narrow
shortest-loop distributions, with large mean values, are the optimal
ones.

In general, we call the emerging structures {\it \bf entangled
networks}: all sites are very much alike (super-homogeneity) and the
links form a very intricate or interwoven structure (no
community-structure, poor modularity, and large shortest-loops). Every
single site is close to any other one (short average distance) owing
not to the existence of intermediate highly connected hubs, as in
scale-free nets
\cite{Reviews}, but as the result of a very ``democratic'' or
entangled structure in which properties such as site-to-site distance,
betweenness, and minimum-loop-size are very homogeneously distributed
(see figs.~1, ~3).

We have tried to use our (so far, partial) understanding of the
entangled-topology to generate them more efficiently. For example, the
constraint of homogeneity in the degree distribution can easily be
implemented by starting up the simulation with regular graphs (or
almost regular graphs) and performing changes respecting such a
property (by randomly selecting pairs of links and exchanging their
endpoints). A much faster convergence to optimal nets is obtained in
this way.  Other topological constraints are not so easy to
implement. We have performed simulations using target functions different
from $Q$ in the optimization algorithm. Functions as the average
distance, average betweenness, or homogeneity measures (such as the
distance variance or the betweenness variance), or $\langle \ell
\rangle$ are not sufficient: they need to be optimized simultaneously,
in some proper way, to obtain reasonable outputs. We have tried
different combinations of these quantities. The best convergence and
results are obtained for the following combination of the betweenness,
$b$, the betweenness variance, $\Delta b$ and $\langle \ell \rangle$:
$U ={((\Delta b)^2+\langle b \rangle^2) \over N} - \langle \ell
\rangle$. The optimization of $U$ is much faster than the minimization
of the eigenratio as $U$ is faster to compute than $Q$. For small
networks the final result is as good as the one of the original method
but, unfortunately, when $N$ increases results worsen, though the
computational time is always relatively small. This failure means that
a full topological understanding of (large) entangled networks has not
been reached yet.

In order to put our findings into context, we discuss some connections
with known concepts in graph-theory. General considerations show that
$\lambda_N \in [k,2 k]$ for regular graphs \cite{Graphs,Sarnak}. As
the variability of $\lambda_N$ is very limited, optimizing $Q$ is
almost equivalent in most cases to maximizing $\lambda_2$ (also called
{\it spectral gap}), as we have verified numerically. It is also known
that for any family of regular graphs, $G_m$ ($m$ is the family
index), in which the size $N$ goes to infinity for large $m$ the
inequality $\lambda_2 \leq k - 2 \sqrt{k-1}$ holds asymptotically,
providing an upper bound for the spectral gap. Finally, it can be
shown that for any family $G_m$ in which the girth goes to infinity
for large $m$, almost all eigenvalues are asymptotically larger or
equal to $k - 2 \sqrt{k-1}$, meaning that the optimal gap value is
typically obtained whenever the girth diverges for large $N$
\cite{Graphs,Sarnak}. These results are in accordance with our
observation of large girths and large $\langle \ell \rangle$ for
entangled networks (even if they are not at the large-$N$ limit).
Another link with graph theory is provided by the concept of {\it
expanders}. These are highly connected sparse graphs, with
applications which include the design of super-efficient communication
networks and de-randomization of random algorithms
 among many others
\cite{Sarnak,Explicit}, and are defined as follows \cite{Sarnak,Graphs}. 
Given a subset $S$ of nodes in a graph $G$, its ``edge boundary'' is
the set of links between nodes in $S$ and nodes in its complement. The
``expansion parameter'' $h$ of a graph $G$
is the minimum ratio of the edge-boundary of a set and the set itself.
A sequence of regular graphs $G_m$ is a \emph{family of expanders } if
its size $N$ diverges for large $m$ and $h$ is always larger than a
given positive constant. This means that the boundary of any subset is
always a non-vanishing fraction of the subset itself. Note that a
large value of $h$ corresponds to a very intricate (entangled)
network, where it is not possible to isolate subsets with a small
boundary (or, in other words, where communities are poorly
defined). Also, the expansion property is strictly related to the
spectral gap
\cite{Sarnak}:
$ \frac{\lam_2}2 \le h(G) \le \sqrt{2 k \lam_2}$, meaning that
(families of) entangled networks are expanders.  {\it Ramanujan
graphs} \cite{Sarnak} are defined as $k$-regular graphs of size $N$
with $ \lambda_2 \geq k - 2 \sqrt{k-1}$. Hence, these graphs are
optimal expanders
\cite{Sarnak}. A family of entangled networks, will typically be a
Ramanujan one (as $\lambda_2$ tends to be maximized) and, therefore, a
(close to optimal) family of expanders. The explicit construction of
expanders and Ramanujan graphs is a currently active field in
graph-theory \cite{Sarnak,Explicit}, and it could serve as a starting
point for explicit entangled-network design.

\begin{figure}
\centerline{ 
\psfig{file=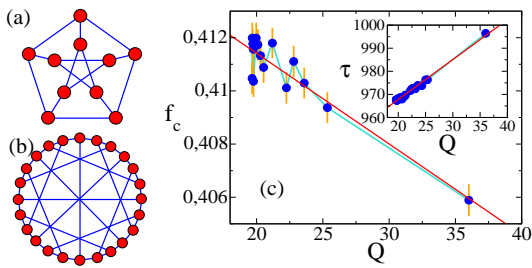,width=7.0cm}}
\caption{Cage graphs for $k=3$ and (a) $g=5$ (Petersen) and (b) $g=7$ (McGee).
Right panel: Percolation threshold (main) and average first-passage
time (inset) as a function of the eigenratio $Q$, as obtained during
the optimization of a network with $500$ nodes and $\langle k\rangle
=3$. The initial network corresponds to a $3$-regular graph with
$N=500$. 
}
\label{Fig3}
\end{figure}

Some properties of entangled networks as related to
other optimization or flow problems follow:

{\bf i)} In a recent paper \cite{2peak}, the optimization of network
robustness against random and/or intentional removal of nodes has been
studied. It was concluded that for generalized random graphs in the
limit $N\to\infty$ the most robust topology (maximizing the
percolation threshold) is characterized by a degree distribution with
no more than $3$ distinct node connectivities; i.e. with a rather
homogeneous degree-distribution. To study the possible connection with
our super-homogeneous entangled networks, let us recall that the
initial topology we have considered (i.e. $k$-regular graphs) is
already the optimal solution for robustness-optimization against
errors and attacks in random networks \cite{2peak}. A natural question
to ask is whether further $Q$-optimization has some effect on the
network robustness.
The answer is yes, as shown in fig. \ref{Fig3} where the percolation
threshold for random or intentional attacks (which coincide for
regular graphs), $f_c$, is plotted versus $Q$ for a particular
$Q$-optimization run. This further improvement of the robustness is
possible because entangled networks include correlations, absent in
random graphs \cite{2peak}.
This tendency is maintained for increasing $N$, confirming that
entangled networks are also extremely efficient from the robustness
point of view (this remains true for {\it reliability} against link
removal \cite{Myrvold}).

{\bf ii)} The problem of optimal topologies for local search with
congestion has been tackled recently \cite{Catalans}, with the
conclusion that when the density of information-packets traveling
through a network is above a given threshold, the optimal topology is
a highly homogeneous one, where all the nodes have essentially the
same degree, betweenness, etc \cite{Catalans}. Again, we encounter
super-homogeneity,
revealing that entangled networks are also
optimal 
for packet flow and local searches with congestion.

{\bf iii)} A typical measure of the network performance for flow
properties is the average first-passage time, $\tau$, of random walks.
It is defined as the average time it takes for a random walker to
arrive for the first time to a given node from another
one. For a $k$-regular graph, $\tau$ can be expressed in terms of the
Laplacian eigenvalues as $\tau \propto \sum
\lambda_n^{-1}$, where the sum runs over all non-zero eigenvalues
\cite{Lovasz}. The largest contribution comes from $1/\lambda_2$,
therefore minimizing $Q$ guarantees a small $\tau$ (see
inset in fig. \ref{Fig3}), providing more evidence that entangled
nets exhibit a very good performance for flow problems.

{\bf iv)} Recently Kim concluded that neural networks with lower
clustering coefficient exhibit much better performance than others
\cite{BJK}. Entangled nets have a very low clustering coefficient as only 
large loops exist and, therefore, they are natural and excellent
candidates to have a good performance and large capacity.

All these features suggest that entangled networks, defined here as
networks which optimize synchronizability, are also extremely good
with respect to many highly desirable properties in networks. This
allows us to state the following \emph{conjecture}: Given $N$ and an
average number of links per site $k$, there exists a network topology
(entangled nets) with many optimal (or almost optimal) features,
characterized by homogeneous degree, betweenness, and distance
distributions, large girths, large average shortest loops, no
community-structure, and small diameters. A more precise topological
characterization of entangled graphs, as well as the definition of an
algorithmic procedure to build them up (similar to those existing for
expanders and Ramanujan graphs), remain open and challenging problems
with a huge amount of potential applications for communication and
technological networks. It seems that these networks do not abound in
the real world; this could be due to the fact that optimal topologies
are not easily reachable within growing network processes. Identifying
examples of these nets and constructing evolution-reachable optimal
networks are fundamental tasks to further gauge the relevance of this
topology in Nature.

\begin{acknowledgments}
  We acknowledge useful discussions with D. Cassi, and
  P. L. Krapivsky, as well as financial support from the spanish MCyT
  (FEDER) under project BFM2001-2841, the MECD, and the EU
  COSIN-project-IST2001-33555.
\end{acknowledgments}


\end{document}